# VIBRATIONAL MEASUREMENT FOR COMMISSIONING SRF ACCELERATOR TEST FACILITY AT FERMILAB*

M.W. McGee[†], J. Leibfritz, A. Martinez, Y. Pischalnikov, W. Schappert,
Fermi National Accelerator Laboratory, Batavia, IL 60510, USA

*Abstract*

The commissioning of two cryomodule components is underway at Fermilab's Superconducting Radio Frequency (SRF) Accelerator Test Facility. The research at this facility supports the next generation high intensity linear accelerators such as the International Linear Collider (ILC), a new high intensity injector (Project X) and other future machines. These components, Cryomodule #1 (CM1) and Capture Cavity II (CC2), which contain 1.3 GHz cavities are connected in series in the beamline and through cryogenic plumbing. Studies regarding characterization of ground motion, technical and cultural noise continue. Mechanical transfer functions between the foundation and critical beamline components have been measured and overall system displacement characterized. Baseline motion measurements given initial operation of cryogenic, vacuum systems and other utilities are considered.

## INTRODUCTION

Cryogenic operation began within the SRF complex during November of 2010. Initial vibration studies helped to characterize technical and cultural noise within the NML environment [1]. Identification of spectral signatures allow for tracking sources within the NML facility. However, many of the rotational equipment sources have the same spectral imprint and dynamic coupling impedes progress towards isolation of these frequencies. These motions are spectrally characterized in terms of root mean square (rms) displacement and Power Spectrum Density (PSD) velocity. In general, ground studies have been conducted at the Fermilab site over the past 15 years, monitoring vibration levels for various projects including: Main Injector, Electron Cooling, Collider Experiments at CDF and D0 and large future colliders [2].

Extensive studies regarding existing cryomodule operation at DESY for the linac of the European X-Ray Free Electron Laser (XFEL) have been completed [3-6]. External ground vibration caused by technical (vacuum, HVAC, water and cryogenic systems) and cultural (staff activity) disturbances manifest fast motion of the focusing quadrupole magnet resulting in pulse-to-pulse beam position jitter [5]. Vertical quadruple motion for cryomodule stability is limited to 30 nm in a frequency range between 10 Hz to 100 Hz. Beam-based feedback through dipole corrector adjustment within the quadrupole package can correct for motion beneath 10 Hz [7].

A 1.3 GHz, type III+ European X-Ray Laser Project (XFEL) cryomodule consists of eight dressed 9-cell niobium superconducting radio frequency (RF) cavities. The cold mass hangs from three column support posts constructed from G-10 fiberglass composite, which are attached to the top of the vacuum vessel. The 312-mm diameter helium gas return pipe (HeGRP), supported by the three columns, acts as the coldmass spine, supporting the cavity string, quadrupole and ancillaries. Brackets with blocks on two sides provide a connection between each cavity and the HeGRP. Two aluminum heat shields (80 K and 5 K) hang from the same two column supports. The coldmass consists of all components found within the 80 K shield shown in Fig. 1. Relative longitudinal and transverse alignment (or position) of the cavity string and quadrupole is held by an Invar rod (a material with very low thermal expansion).

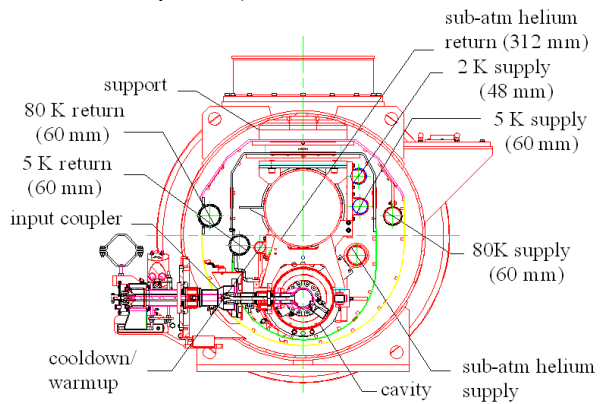

Figure 1: CM1 cryomodule section view.

## OPERATIONAL MEASUREMENT

A seismic station located on the floor of NML near the center of CM1 was established to help define external input such as ground motion. Within the cryomodule CM1 geophone instrumentation was mounted on the cavities and quad.

### Instrumentation

The cryomodule CM1 was instrumented with Oyo Geospace GS-14-L9 geophones [8] mounted on a supporting bracket above cavity #3 (Cav3), #5 (Cav5) and #7 (Cav7), one vertical and one horizontal. These cryomodule CM1 cavity geophones were enclosed within a cylindrical magnetic shield made of heat-treated 1018 low-carbon steel to protect the cavities from stray

---

magnetic fields. A vertical Oyo Geospace GS-11-D geophone [8] and horizontal geophone SM6-HB from Sensor b.V [9] were attached on the quad, upstream. These geophone devices were used within cryomodule CM1 successfully down to 2 K without a loss of sensitivity. Cold calibration of these geophones has been established at DESY [5]. Inertial (moving coil) velocity sensors Geospace HS-1 (2 Hz) geophone [8] devices in tri-axial aluminum block sets were used with vertical Sercel Mark L4c seismometers [10] as a baseline within the NML seismic station. Each device was connected to a National Instruments (NI) NI-9233 4-channel, 24-bit ADC modules sampled at 1,500 K/sec, and the data was collected using a dedicated PC desktop.

*Transfer Function Measurements*

The vibration level measurement during cryogenic/RF tests on the CryoModule Test Bench (CMTB) at DESY under room temperature conditions revealed that rigid body motion occurred between the vacuum vessel and the quad (coldmass). These results indicated no mechanical resonance occurs below 50 Hz and 80 Hz in the transverse and vertical direction, respectively [6]. The rigid body motion response is an inherit property of the Type III+ cryomodule design, and therefore cryomodule CM1 shares a similar response.

Isolation of vacuum pumping sources was necessary to confirm spectral responses to cultural and technical noise within the NML Facility. Ground motion between 1 to 100 Hz typically peaked at 0.03 microns rms due to limited local traffic within the Fermilab site and technical noise within NML. The effect of local traffic was insignificant since NML is fairly remote and the floor exists 5 m below grade level. Mechanical stability of the system was considered through transfer function amplitude measurements between cryomodule CM1 internal components and the floor (shown in Fig. 2) revealed resonant responses above 55 Hz and 61 Hz in the transverse and vertical direction, respectively.

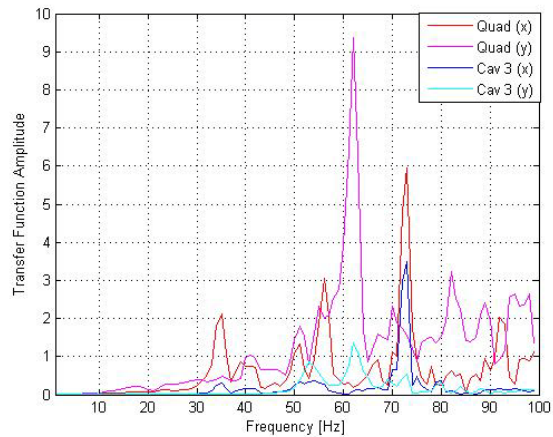

Figure 2: Vertical (y) and transverse (x) transfer function between cryomodule CM1 coldmass devices and floor.

*Liquid Helium Operation*

The cooldown began in mid November of 2010 as turbulent gas flow of initial circulation of cold nitrogen and helium gas within the cryomodule CM1 piping circuit. The first measure of liquid helium at 4.2 K formed two days later, shown in Fig. 3 within the cavity string of cryomodule CM1 occurred on November 19[th] of 2010. Erratic helium gas flow with random pressure spikes caused the cavities, quad and HeGRP to resonate. This behavior quickly became quiescent as liquid helium began to form and fill each cavity.

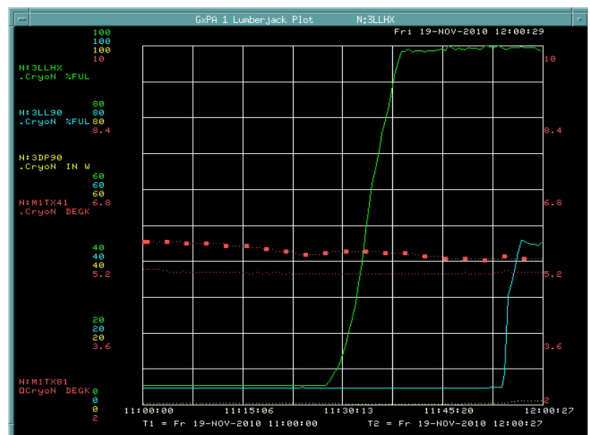

Figure 3: ACNET (controls) plot of final cooldown phase to 4.2 K (liquid fill).

*Superfluid Operation*

Superfluid helium within the cryomodule CM1 cavity string was achieved on November 22[nd], 2010. Vertical and horizontal vibration measurements taken during operation at 2 K (23.4 Torr) reflect the stability of cryomodule CM1 without RF power. Vertical quad integrated displacement rms during two hours of operation given in Fig. 4 show a range of motion between 0.47 and 0.65 microns at ~19 Hz, 0.35 to 0.47 microns at 25 Hz and 0.2 to 0.35 microns at 30 Hz. The cavities: Cav3, Cav5 and Cav7 have a similar response, however at lower displacements. A significant portion of cavity Cav3 displacement occurred at 25 Hz, ranging from 0.09 to 0.3 microns.

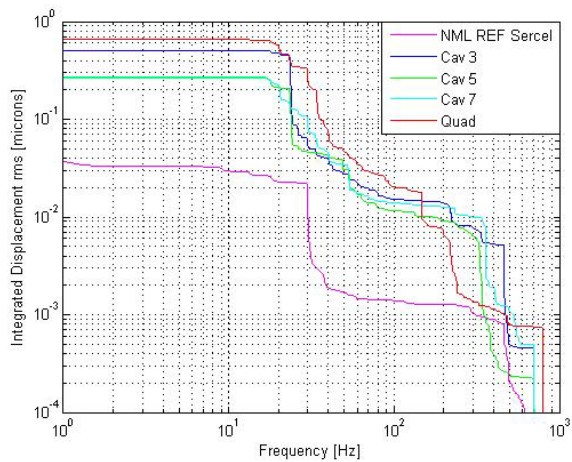

Figure 4: Integrated vertical displacement for superfluid operation.

Transverse quad integrated displacement rms in Fig. 5 ranged from 0.22 to 0.49 microns at ~18 Hz. This was slightly higher in frequency than the predicted ANSYS finite element (FE) transverse pendulum (or rocking) mode of 11.2 Hz [7]. At 25 Hz and 30 Hz, transverse motion ranging from 0.15 to 0.22 microns and 0.07 to 0.15 microns occurs, respectively. This motion was attributed to the rotational equipment such as vacuum pumps and other utilities with speeds ranging between 1,500 to 1,800 rpm. Documented site specific frequencies of 4.2, 9.2 and 13.2 Hz have been observed within the Fermilab site due to the Central Helium Liquefier (CHL) [11], located 2.4 kilometers from NML. The 4.2 Hz frequency was seen in the ground motion; however it was not visible at the coldmass level.

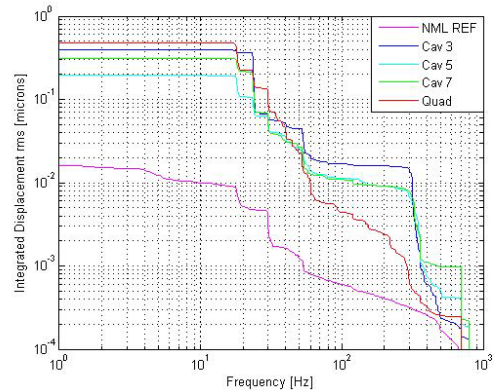

Figure 5: Integrated horizontal displacement for superfluid operation.

*Flow Measurement*

Liquid helium flow rate measurements were taken from a static regime up to 6 g/sec using a FCI Model ST98. This work involved checking the accuracy of the external flow meter located on the vacuum pump exhaust. Tests were performed without RF power after one day of stability while operating at a certain heater power. Integrated displacement rms data for each component was constant, based on a one hour integration time prior to flow rate measurement for a given heater power setting.

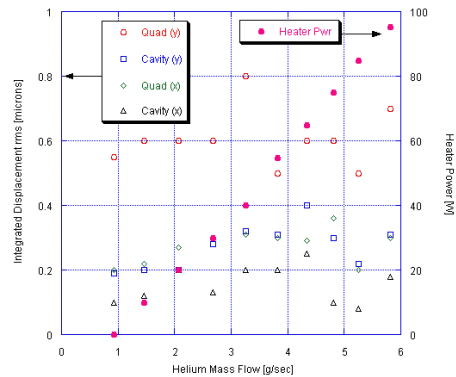

Figure 6: Operational 2K flow rate studies versus transverse (x) and vertical (y) quad and cavity motion.

## FUTURE WORK

Cavity RF power studies continue as cavities are individually tested. Evidence of Thermal Acoustic Oscillation (TAO) has been found during peak power dissipation events into the liquid helium. The use of vertical displacement rms is being implemented as an operational flag regarding cavity instability through the control system ACNET. Cavity and quad geophones continue to provide a useful diagnostic tool during cryogenic commissioning and stable operation.

## ACKNOWLEDGEMENTS

We wish to thank Mack Amorn-vichet, Ryan Heath, Stewart Mitchell, Tony Parker, Dave Slimmer and Nino Strothman (Computer, Networking and Labview Application Support Personnel). Also, thanks to Brian DeGraff, Greg Johnson and Arkadiy Klebaner for cryogenic support. Special thanks to Kermit Carlson and Wayne Johnson for their technical support.